# On Source Rules for ABR Service on ATM Networks with Satellite Links*

Sonia Fahmy, Raj Jain, Shivkumar Kalyanaraman, Rohit Goyal, and Fang Lu†

Department of Computer and Information Science
The Ohio State University
Columbus, OH 43210-1277, USA
Email: {*fahmy, jain, shivkuma, goyal, flu*}@cis.ohio-state.edu

**Abstract** During the design of ABR traffic management at the ATM Forum, we performed several analyses to ensure that the ABR service will operate efficiently over satellite links. In the cases where the performance was unacceptable, we suggested modifications to the traffic management specifications. This paper describes one such issue related to the count of missing resource management cells (Crm) parameter of the ABR source behavior. The analysis presented here led to the changes which are now part of the ATM traffic management (TM 4.0) specification. In particular, the size of the transient buffer exposure (TBE) parameter was set to 24 bits, and no size was enforced for the Crm parameter. This simple change improved the throughput over OC-3 satellite links from 45 Mbps to 140 Mbps.

## 1 Introduction

Satellite communication systems are the means of realizing a worldwide Broadband Integrated Services Digital Network. Satellites play a crucial role in the integration of networks of various types and services, and will potentially be used for a wide range of applications. Satellites will continue to play an ever-increasing role in the future of long-range communications [10].

The long propagation delay of the geosynchronous earth orbit (GEO) satellite links imposes severe demands and constraints on the networking protocols that can be employed in a satellite network. It is extremely important to verify that networking protocols can operate efficiently on such long-delay links. In this paper, we investigate the operation of the Asynchronous Transfer Mode (ATM) Available Bit Rate (ABR) service over satellite links. We explore the scalability of the ABR source end system behavior and parameter values to satellite networks, and point out important issues through analysis and simulation results.

Asynchronous Transfer Mode (ATM) is proposed to transport a wide variety of services, such as voice, video and data, in a seamless manner. ATM cells flow along predetermined paths called Virtual Circuits (VCs). End systems must set up Constant Bit Rate (CBR), Variable Bit Rate (VBR), Available Bit Rate (ABR) or Unspecified Bit Rate (UBR) VCs prior to transmitting information.

Data traffic in ATM is expected to be transported by the ABR service. The ATM Forum Traffic Management group has standardized a rate-based flow control model for this class of traffic. The ABR flow control standards specify the behavior of the source and destination end systems, and establish guidelines for the operation of ATM switches.

†Fang Lu is currently with Lucent Technologies, NJ.

The rules specifying the behavior of the source and destination end systems have undergone major revisions throughout the past two years. We have actively participated in that revision process. This paper examines the analysis and results that led to a change to source end system rule 6 parameters to allow efficient operation over satellite links. An overview of end system behavior and an explanation of rule 6 are presented next, followed by the analysis of the effect of rule 6 on satellite networks.

# 2 ABR Source End System Behavior

The ABR source end system is allowed to send data at a given rate called "Allowed Cell Rate" (ACR), which ranges between a negotiated Peak Cell Rate (PCR) and Minimum Cell Rate (MCR). Immediately after establishing a connection, ACR is set to an Initial Cell Rate (ICR), which is also negotiated with the network. The source sends a Resource Management (RM) cell every Nrm−1 data cells (default Nrm value is 32), and the destination end system turns the RM cells around. The RM cells traveling from the source to the destination are called forward RM cells, while the RM cells traveling from the destination back to the source are called backward RM cells. The RM cells collect the network feedback (which is based on the network load) and are returned to the source, which adjusts its allowed cell rate according to that feedback.

Most RM cells are generated by the source. However, if a switch is extremely congested, it can generate an RM cell and send it backward towards the source without waiting for the returning RM cells. This is called backward explicit congestion notification (BECN). RM cells contain a flag called "backward notification" (BN), which indicates whether the RM cell was generated by the source (BN=0) or the switch (BN=1). Receipt of backward RM cells with BN=0 at the source indicates that there is no physical discontinuity in the round trip path.

A complete explanation of the thirteen source rules is presented in [6]. Here we concentrate on source rule 6, which is intended to protect the source from losing packets due to broken links on the path. The rule requires the sources to reduce the rate if no feedback is received from the network.

## 2.1 Source Rule 6

In the ATM Forum Traffic Management specifications [2], source end system rule 6 is stated as:

*"Before sending an in-rate forward RM cell, and after adjusting ACR according to number 5 above, if at least Crm in-rate forward RM cells have been sent since the last backward RM cell with BN=0 was received, then ACR shall be reduced by at least ACR ×CDF, unless that reduction would result in a rate below MCR, in which case ACR shall be set to MCR"*

Basically, the sources keep a count of the RM cells sent, and if no backward RM cells are received for a long time, the sources reduce their rate by a factor of "Cutoff Decrease Factor (CDF)" usually in the range of 1/64 to 1. The "long time" is defined as the time to send Crm forward RM cells at the current rate. Here, Crm is a parameter.

During connection setup, the sources negotiate several parameters with the network. One such parameter is "transient buffer exposure (TBE)." The name comes from the fact that this parameter determines the exposure of the switch to sudden traffic transients. It determines the the number of cells that may be received at the switch during initial start up (or after any long idle period of time). After TBE cells, the source reduces its input rate as described above. TBE is specified in cells while Crm is specified in RM cells. Since there is one RM cell per Nrm cells, the relationship between

Crm and TBE is as follows:

$$\text{Crm} \leftarrow \lceil \text{TBE}/\text{Nrm} \rceil$$

The next section explores the operation of source rule 6 and its parameters in more depth.

## 2.2 Operation of Source Rule 6

While source rule 6 was designed to limit the number of cells lost if any link along the path breaks, it also triggers if the network is highly congested. In particular, it triggers if the rate of backward RM cells returning from the network is less than (1/Crm)th of the rate of forward RM cells entering the network. This can be understood from figure 1, which illustrates the flow of RM cells in the network.

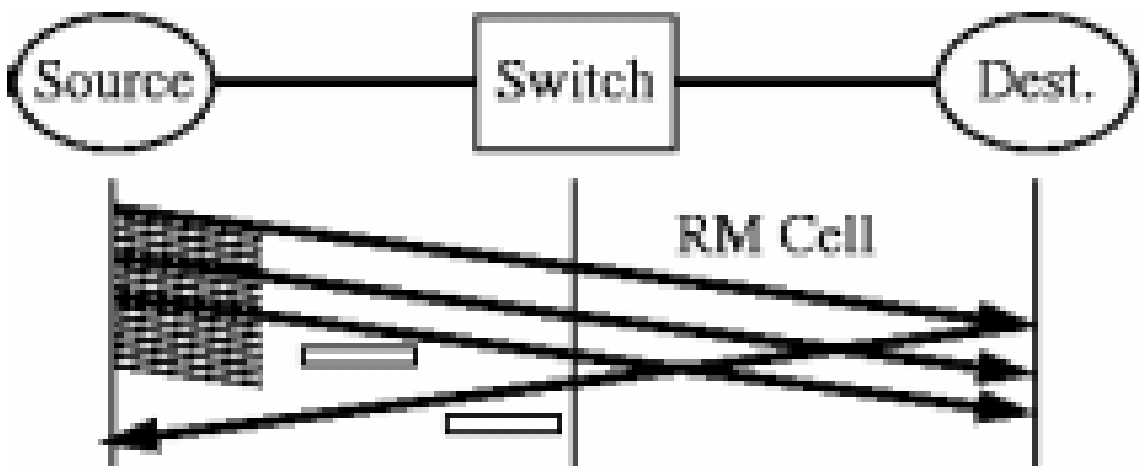

Figure 1: RM cell flow in the network.

Assume that a certain source $S$ is sending forward RM cells at an average rate of $R$ cells per second (cps). The RM cells are turned around by the destination and the backward RM cells are received by $S$ at a different rate $r$ cps. In this case, the inter-forward-RM cell time at the source is $1/R$ while the inter-backward-RM cell time at the source is $1/r$. Source end system Rule 6 will trigger at $S$ if the inter-backward-RM time is much larger (more than Crm times larger) than the inter-forward-RM time [7]. That is, if:

$$1/r \geq \text{Crm} \times (1/R)$$

or:

$$R \geq \text{Crm} \times r$$

When the source rule 6 is triggered, the source reduces its rate by a factor of CDF but not below the minimum cell rate. That is,

$$\text{ACR} \leftarrow max(\text{MCR}, \text{ACR} - \text{ACR} \times \text{CDF})$$

where the value of CDF can be zero (for no rate decrease), or it can be a power of two that ranges from 1/64 to 1.

This means that after Crm RM cells are sent (or Crm×Nrm total cells are sent), and no backward RM cell is received:

$$\text{ACR} = \text{ACR}_{initial} \times (1 - \text{CDF})$$

Note that if rule 6 is triggered once, it usually triggers on sending successive forward RM cells (as long as no backward RM cells are being received).

Thus, after Crm+1 RM cells (or (Crm+1)×Nrm cells) are sent:

$$\text{ACR} = \text{ACR}_{initial} \times (1 - \text{CDF})^2$$

After Crm+$k$ RM cells (or (Crm+$k$)×Nrm cells) are sent:

$$\text{ACR} = \text{ACR}_{initial} \times (1 - \text{CDF})^{k+1}$$

Such repeated rate reductions result in an exponential rate drop when source rule 6 triggers, as long as no feedback is being received as shown in figure 2.

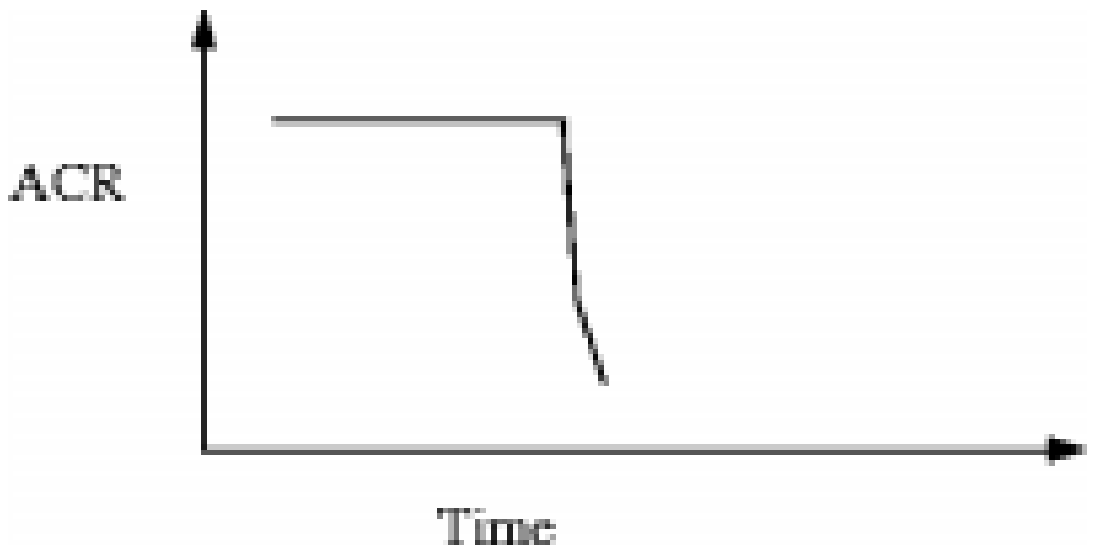

Figure 2: Rule 6 results in a sudden drop of rate.

The effect of source rule 6 on ATM networks with satellite links is examined next.

# 3 Effect of Rule 6 on Satellite Links

For long-delay links, such as satellite links, our simulation results revealed that source rule 6

can unnecessarily trigger and cause oscillations during start up and after idle periods. This can degrade the throughput considerably. In this section, we present simulation results showing the problem then we present analytical arguments and a solution. We simulated several different configurations. However, the problem was common to all configurations. Therefore, here we present only the simplest configuration consisting of a single VC going through a single satellite link.

### 3.1 Simulation Parameters

Figure 3 shows the "one source configuration" used to illustrate the problem.

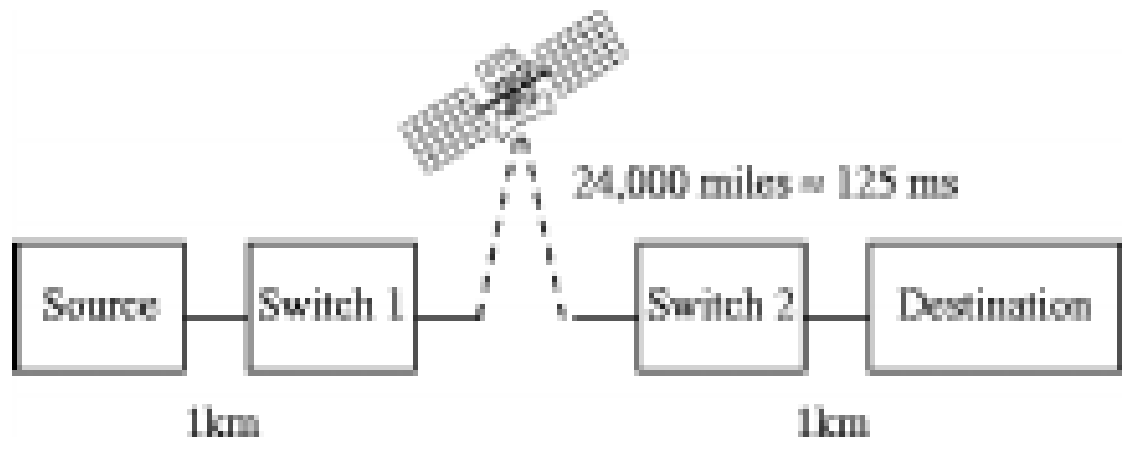

Figure 3: One source configuration.

All the links are OC-3 links operating at a rate of 155.52 Mbps. The link connecting the two switches is a satellite link, while the links connecting the switches to the end systems are each 1 Km long (Local Area Network (LAN) links). The one-way propagation delay of the satellite link is 275 ms, while the propagation delay of each LAN link is 5 microseconds.

The traffic is bidirectional, and the sources are persistent (sources always have data to send). The ABR source parameter values are listed in table 1.

The switches employ the ERICA switch scheme [8]. The ERICA algorithm uses two key parameters: target utilization and averaging interval length. The algorithm measures the average load and number of active sources over successive intervals and attempts to achieve a link utilization equal to the target. The averaging intervals end either after a specified period of time, or after a specified number of cells have been received, whichever happens first. In the simulations reported here, the target utilization is set at 90%, and the averaging interval length is set to the minimum of 30 ABR input cells and 20 microseconds. The small averaging interval value provides the best performance for the infinite traffic case studied here.

### 3.2 System Performance

Before August 1995, Crm was an 8-bit integer with a default value of 32. Figure 4 illustrates the performance of the system with Crm set to 32. Figure 4(a) shows the allowed cell rate of the source over 1200 ms, and figure 4(b) shows the number of cells received at the destination during the same period of time.

As seen in figure 4(a), the initial rate is 140 Mbps (90% of 155 Mbps). After sending 32 RM cells (or Crm$\times$Nrm $= 32\times32 = 1024$ cells), rule 6 triggers and the rate rapidly drops. The first feedback is received from the network after around 550 ms (275 ms$\times$2), because the one-way delay of the satellite link is 275 ms. The network asks the source to go up to 140 Mbps. The source increases its rate but rule 6 triggers again. The rule triggers again because the time between returning RM cells is large (they were sent at a low rate). Due to rule 6, the source reduces its rate.

This phenomenon of increase and decrease repeats resulting in high-frequency oscillations between very low rates and very high rates. The rapid rate drops occur due to the triggering of source rule 6, while the rate increases occur because the network feedback is consistently at 140 Mbps (90% of 155 Mbps) [1].

Figure 4(b) shows the number of cells received at the destination. From this figure, it is possible to compute instantaneous throughput by computing the slope of the curve. It is also possible to compute average throughput over any interval by dividing the cells received (increase in the $y$-value) during that interval by the period of time ($x$-value) of the interval. The average throughput during the interval from 275 ms

Table 1: ABR Source Parameters

| Parameter | Explanation | Value |
|---|---|---|
| PCR | Peak Cell Rate | 155.52 Mbps |
| MCR | Minimum Cell Rate | 0 Mbps |
| ICR | Initial Cell Rate | 0.9×PCR = 140 Mbps |
| Nrm | Every $\text{Nrm}^{th}$ cell is an RM cell | 32 |
| RIF | Rate Increase Factor | 1 |
| CDF | Cutoff Decrease Factor | 1/16 |
| Crm | Count of missing RM cells | 32, 256, 1024, 4096, 6144, 8192 |

to 825 ms is 32 Mbps and that during the interval from 825 ms to 1200 ms is 45 Mbps.

During the first 550 ms, the source is mostly sending at a very low rate until the first feedback is received after about 550 ms. The effect of the receipt of feedback can be observed at the destination after 550+275=825 ms. After the first feedback is received, the rate oscillations result in reduced throughput. The results do not significantly vary for different values of CDF.

## 3.3 Analysis

The low throughput values in figure 4(b) are a result of the unnecessary triggering of source rule 6 for small Crm values. Rule 6 limits the number of cells that can be in flight during start up periods.

For a Crm value of 256 (which was the maximum allowed value when Crm was an 8-bit integer before August 1995), there can be 8192 cells in flight, but this is still insignificant compared to the number of cells that can be in flight in case of satellite links. Due to the long propagation delay of the satellite links, the round trip time is longer, and hence the number of cells that can be in flight is larger.

For full throughput, we need to set the value of Crm such that the number of cells in flight can be as large as those required to fill the path both ways. This number is equal to the round trip time (RTT) multiplied by the link rate. Hence, the number of RM cells in flight (Crm) should be (1/Nrm)th of this value:

$$\text{Crm} \geq \text{RTT} \times \text{Link Bandwidth}/\text{Nrm}$$

For 155 Mbps links, Crm should be greater than or equal to 6144 (550 ms×365 cells per ms/32 cells).

Figure 5 shows the simulations results for a Crm value of 6144. Observe that the allowed cell rate remains at its optimal value and the throughput is high.

For 622 Mbps links, Crm should be greater than or equal to 24576 (6144×4).

For two 622 Mbps satellite hops, Crm should be greater than or equal to 49152 (24576×2).

For $n$ 622 Mbps satellite hops, Crm should be greater than or equal to $24576\times n$.

These numbers indicate that 8 bits are not sufficient to store the required value of the Crm parameter for long round trip delays or high speeds. For example, 49152 requires 16 bits of storage. This analysis resulted in setting the size of the TBE parameter to 24 bits. Since Nrm is normally 32, a 24-bit TBE allows a 19-bit Crm, which is sufficient for most situations.

Initially, the specification recommended that Crm be stored in an 8-bit register. We succeeded in removing this recommendation. Since Crm is an internal parameter that is not negotiated, there is no need to recommend or enforce any particular size.

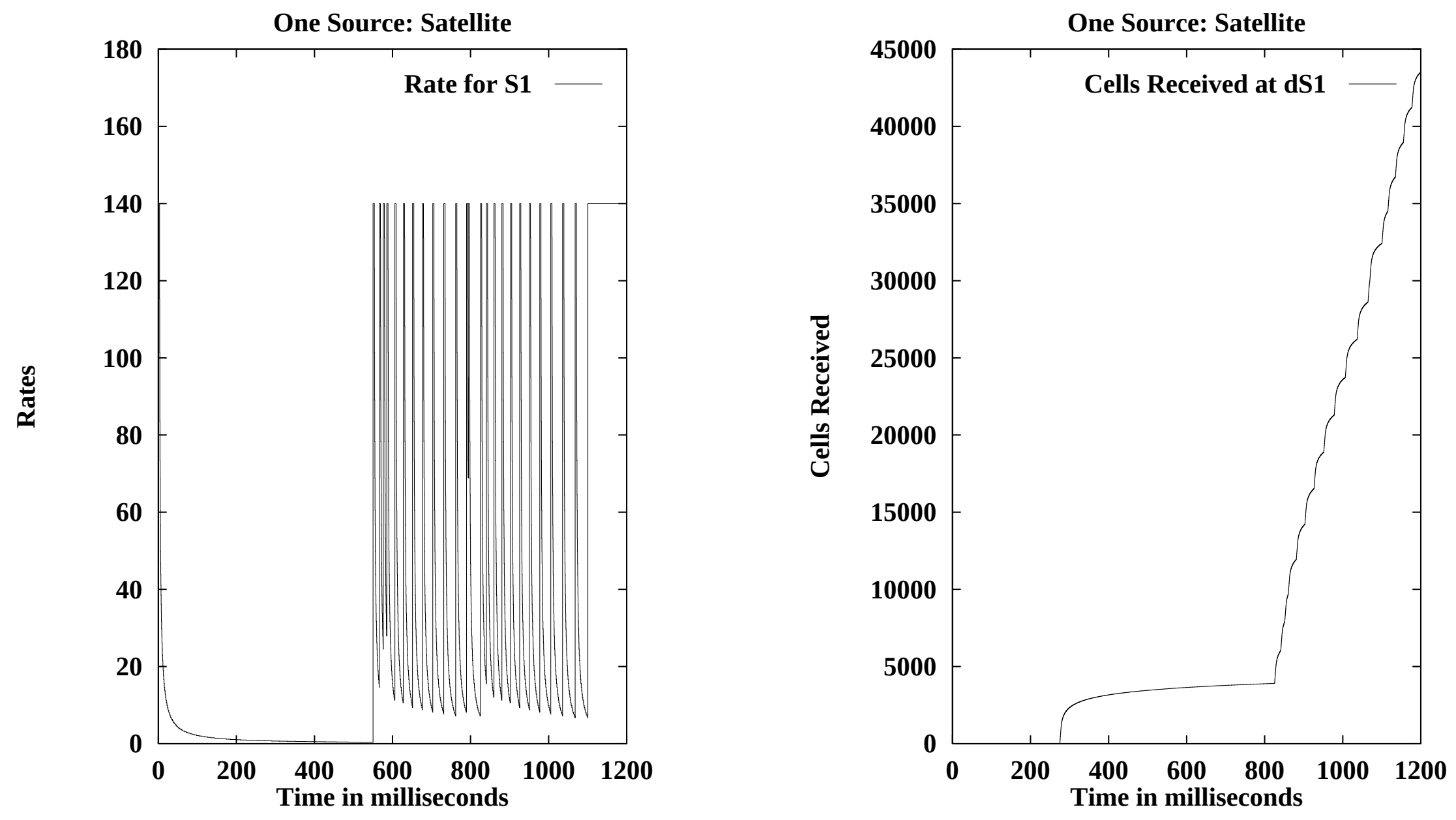

(a) Allowed Cell Rate (ACR)

(b) Cells Received at Destination

Figure 4: Simulation results for a one source configuration. Crm = 32.

### 3.4 Alternatives

The Crm debate continued at the ATM Forum for at least three successive meetings. Some people argued that a large Crm will require too many bits to store it in the network interface cards (NICs). Since a NIC may support several thousand VCs, each extra bit results in several thousand more bits of memory on the board. The problem with this approach is that those NICs that implement short Crm will not be able to efficiently operate over satellite links.

If the specification recommended an 8-bit Crm register, most NIC vendors will implement only 8-bits and most NICs will not be able to operate efficiently over satellite links. Users of satellite links will have to buy special NICs with larger Crm registers.

It is more convenient if the buyers do not have to worry about the links and any user can talk to any other user regardless of the path. (This is how it works on the telephone links today. We do not need special phones to connect over satellite paths). We, therefore, argued in favor of removing the 8-bit recommendation.

## 4 Conclusions

As a general design principle, abnormal operation should not be handled at extreme cost to normal operation. While we do not want to lose too many cells if the link breaks, we do not want to get 50% throughput if the link is operating normally.

In addition to this, if the network is operating optimally, the control scheme should not move it to suboptimal operation. If the source is sending at the optimal rate, its rate should not be changed.

In order for the above principles to be applied in satellite networks, it was necessary to remove the restriction on the size of the Crm ABR source parameter. Both analysis and simulation results confirmed that large values of Crm are necessary for long delay links or high speeds. The motion requesting that "Crm is an integer whose size is implementation depen-

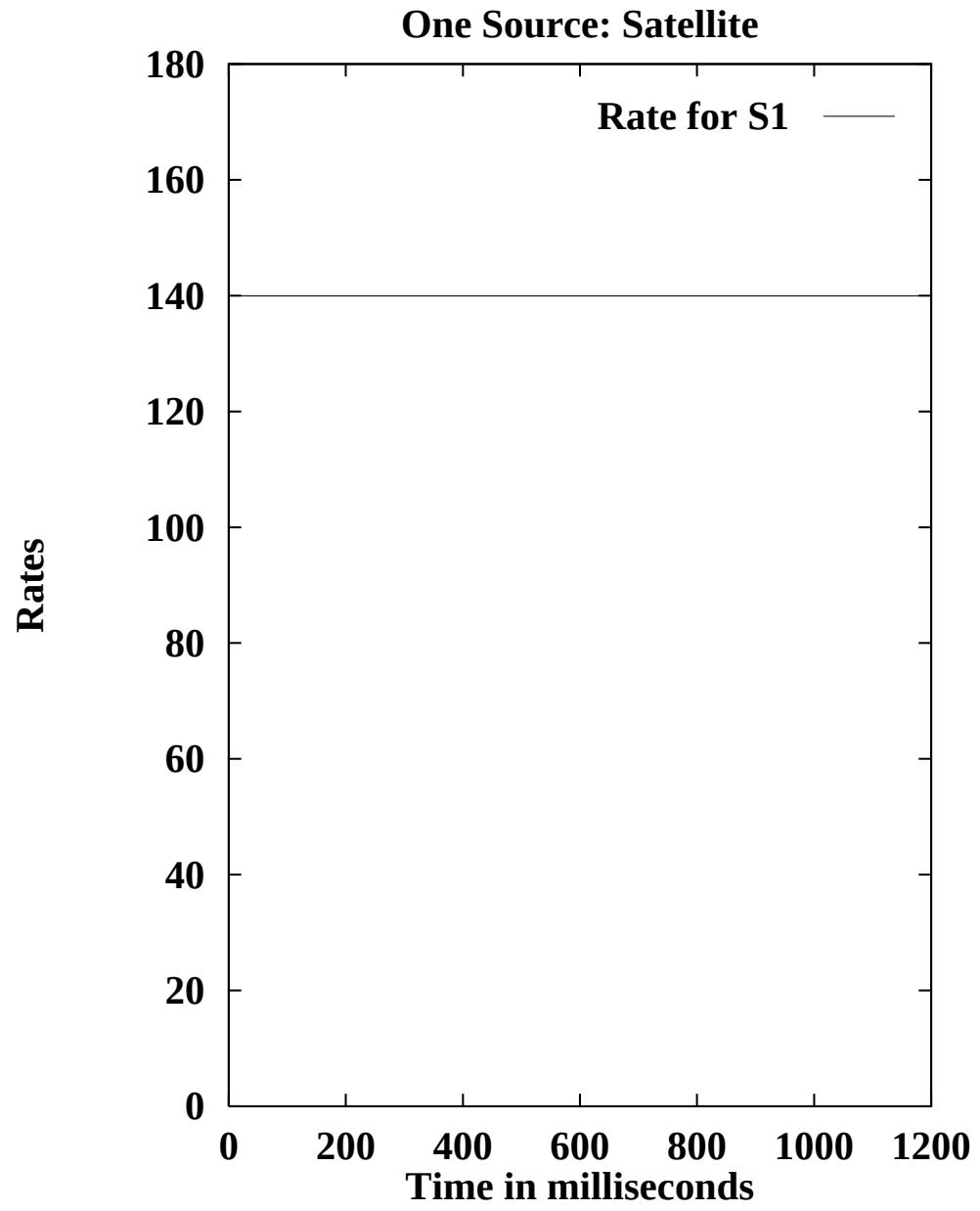

(a) Allowed Cell Rate (ACR)

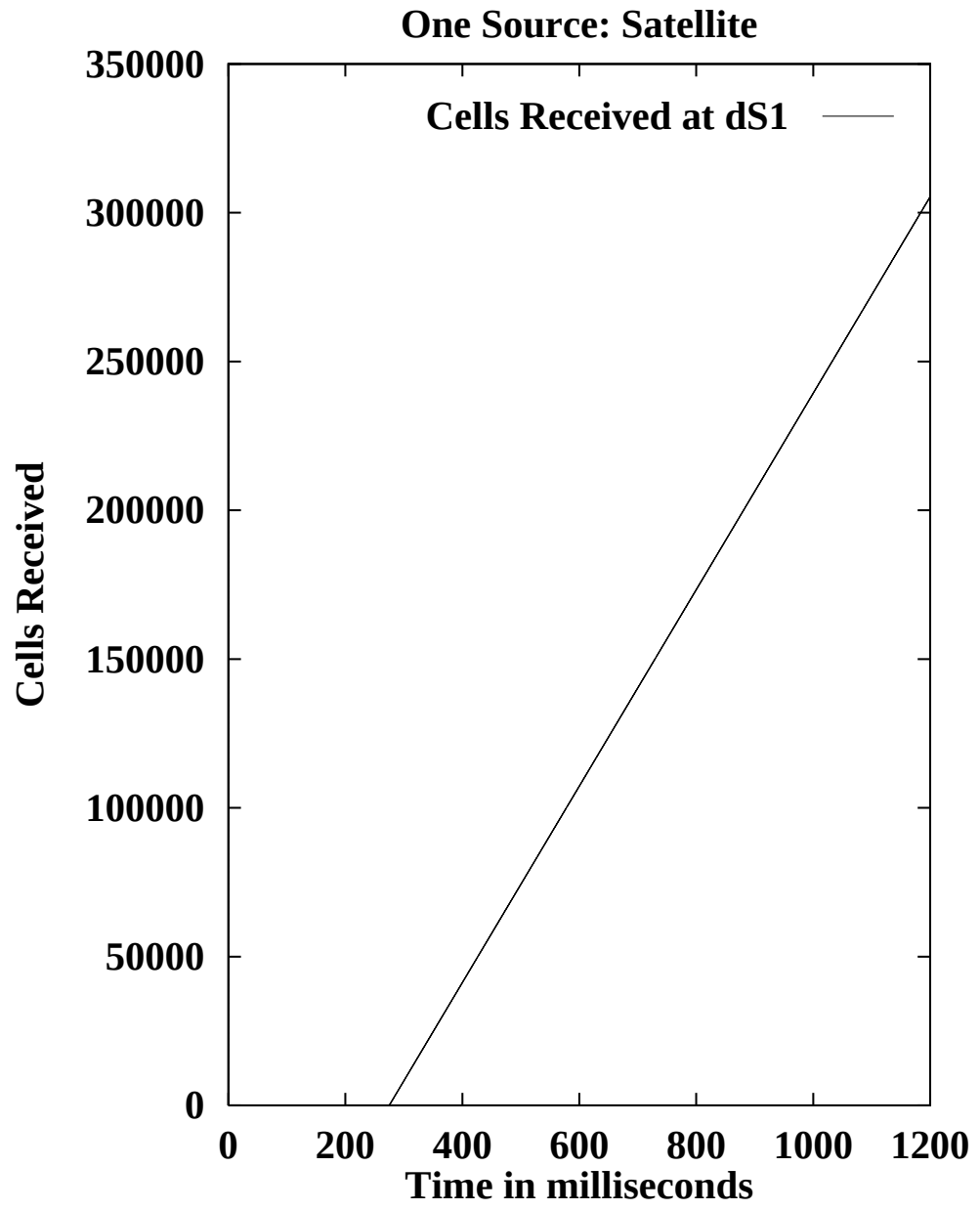

(b) Cells Received at Destination

Figure 5: Simulation results for a one source configuration. Crm = 6144.

dent" has been passed by the ATM Forum Traffic Management working group, and is included in the current specifications.

[1] All our papers and ATM Forum contributions are available through http://www.cis.ohio-state.edu/~jain